\documentclass[11pt,fleqn]{article}

\usepackage{amsmath,amsfonts,amssymb,latexsym}

\usepackage{graphicx}
\usepackage{mathptmx}
\usepackage{bm}
\usepackage{amssymb}
\usepackage{amsmath}
{\hspace*{\fill}{\protect\small {\bf Bijan~Saha}} \hspace*{\fill} }
{\hspace*{\fill} {\protect\small {\bf Nonlinear spinor field with Lyra’s geometry: Bianchi type-VI space-time}} \hspace*{\fill} } 

\newcommand {\cG}{\cal G}
\newcommand {\cD}{\cal D}
\newcommand {\bg}{\bar \gamma}
\newcommand {\bp}{\bar \psi}

\newcommand {\G}{\Gamma}

\def\myfigure #1#2#3#4
{\begin{figure}[ht]\begin{center}
			\includegraphics[width=#2 \textwidth]{#1.png}
			\parbox[t]{#4\textwidth}{\caption{#3}\label{#1}}
\end{center}\end{figure}}

\def \myfigures #1#2#3#4#5#6#7#8
{\begin{figure}[ht]
		\begin{center}
			\includegraphics[width=#2 \textwidth]{#1.jpg}
			\hfill
			\includegraphics[width=#6 \textwidth]{#5.jpg}
			\parbox[t]{#4\textwidth}{\caption {#3}\label{#1}}
			\hfill
			\parbox[t]{#8\textwidth}{\caption {#7}\label{#5}}
		\end{center}
\end{figure} }

\begin{document}
	
	\baselineskip -24pt
	\begin{center}
		\title{Nonlinear spinor field with Lyra’s geometry: Bianchi type-VI space-time}
		{\bf Bijan Saha}\\
		
		{Laboratory of Information Technologies\\
			Joint Institute for Nuclear Research\\
			141980 Dubna, Moscow region, Russia\\ and\\
			Peoples' Friendship University of Russia (RUDN University)\\
			6 Miklukho-Maklaya Street, Moscow, Russian Federation\\ and \\
			Dubna State University\\
			Dubna, Moscow reg. Russia\\ \vskip 1 mm
			orcid: 0000-0003-2812-8930 }\\ \vskip 1 mm
		email:{bijan@jinr.ru}\\
		URL: {http://spinor.bijansaha.ru}
	\end{center}

	\hskip 1 cm

	\begin{abstract}
	In the context of a Bianchi type-VI space-time characterized by Lyra’s geometry, we investigate the influence of a nonlinear spinor field on the evolution of the Universe. Our previous research has examined the nonlinear spinor field within Bianchi diagonal models, revealing that the spinor field exhibits non-trivial non-diagonal components of the energy-momentum tensor. These components impose various constraints on both the space-time geometry and the spinor field itself. The incorporation of Lyra’s geometry into the framework does not alleviate these constraints; however, it alters the preservation of the energy-momentum tensor. Furthermore, this integration complicates the relationship between the invariants of the spinor field and the space-time, ultimately affecting the outcomes of our analysis. Moreover, in this case energy-momentum tensor does not preserve. 
	\end{abstract}
	
	keywords: {spinor field; BVI cosmology; Lyra's Geometry}
	
	pacs: {98.80.Cq}
	
	%\maketitle
	
	\bigskip
	
	\section{Introduction}
	
	Shortly after Einstein proposed his famous theory of gravity, Weyl 
	in an attempt to unify gravitation and electromagnetic field, introduced a generalization of Riemannian Geometry \cite{Weyl}. Weyl theory was not taken seriously as it contradicted some well-known observational result. In 1951 Lyra proposed a modification of Riemannian geometry which bears a close resemblance of Weyl geometry \cite{Lyra}. But unlike Weyl geometry, in Lyra's geometry the connection is metric preserving as in Riemannian geometry. In doing so he introduced a gauge function into the structure-less manifold. This theory was further developed by Scheibe \cite{Scheibe}, Sen \cite{Sen1957}, Halford  \cite{Halford}, Sen and Dunn \cite{Sen1971}, Sen and Vanstone \cite{Sen1972} and many others. Recently Lyra's geometry is being used extensively in cosmology \cite{Soleng,Beesham,Jahromi,Bakry,Shchigolev}.

	 In recent years, many authors considered spinor field as an alternative source  
	 of gravitational field which is able to explain a number of problems in cosmology such as 
	 initial singularity, late time acceleration etc. 
	 \cite{Saha1997GRG,Saha1997JMP,SahaPRD2001,Greene,ELKO,PopPRD}.
	 Several studies show that the presence of nontrivial, non-diagonal
	 components of its energy-momentum tensor (EMT) independent to spinor field nonlinearity 
	 impose severe constraints on both the space-time geometry and the nonlinearity of
	 the spinor field itself \cite{Saha2016,Saha2018}. The spinor field
	 within Lyra's geometry has also been explored 
	 \cite{Casana,SahaJMP}. Recently we have studied a nonlinear spinor field in  
	   a Bianchi type-I anisotropic space-time with Lyra's geometry \cite{SahaJMP}. The goal of this paper is to further examine the role of Lyra geometry when the universe is filled with spinor field. As a study case we consider a Bianchi type-VI gravitational cosmological model.

\section{Basic equations} 

	An affine connection is characterized by its components $\Gamma^\mu_{\alpha \beta}$ which are defined by the change due to infinitesimal parallel transform 
	of a vector  $\xi^\mu$ from a point  $P(x^\mu)$ to a point $P(x^\mu + dx^\mu)$:

		\begin{align}
			\delta \xi^\mu = - \Gamma^\mu_{\alpha \beta} \xi^\alpha dx^\beta, \label{connec} 
		\end{align}
	
and the fundamental metric tensor $g_{\mu\nu}$ that is defined the measure of length $\xi$ of a vector $\xi^\mu$:
		\begin{align}
			\xi^2 = g_{\mu\nu} \xi^\mu \xi^\nu. \label{connec1} 
		\end{align}
 In Riemannian geometry torsion remains absent, i.e., $\Gamma^\alpha_{\mu\nu} = \Gamma^\alpha_{\nu\mu} $ and under parallel transform length of a vector remains unaltered, i.e., $\delta\xi = 0$. Variation of \eqref{connec1} in view of \eqref{connec} gives expression for connection $\Gamma^\mu_{\alpha \beta}$. In Riemannian geometry $\Gamma^\mu_{\alpha \beta}$, known as the Levi-Civita connection, is found to be
  \begin{align}
 	\Gamma^{\mu}_{\alpha \beta} = \frac{1}{2} g^{\mu \nu}\left(\partial_\alpha g_{\nu\beta} + \partial_\beta g_{\alpha\nu} - \partial_\nu g_{\alpha\beta} \right). \label{L-C}
 \end{align} 
	In Riemannian geometry $g_{\mu\nu; \alpha} = 0.$

\subsection{Weyl's geometry:}

 According to Weyl, there exist a geodesic gauge in which length of a vector does not change under parallel transform, but in an arbitrary gauge it is assumed to change

\begin{align}
	d\xi  = - \xi \phi_\mu dx^\mu, \label{arbgau} 
\end{align}
where $\phi_\mu$ is a vector function characterizing the manifold. Thus the metrical connection of a Weyl manifold is characterized by two independent quantities $g_{\mu\nu}$ and 
$\phi_\mu$. If one makes a gauge transformation $\xi \to \bar\xi = \lambda \xi$, where $\lambda = \lambda (x)$ then
$g_{\mu\nu}$ and $\phi_\mu$ transforms as follows \cite{Sen1971}
\begin{align}
	g_{\mu\nu} \to  \bar g_{\mu\nu} = \lambda g_{\mu\nu}, \quad \phi_\mu = \phi_\mu - \lambda_\mu/\lambda, \quad \lambda_\mu = \partial\lambda/\partial x^\mu. \label{gaugeWeyl} 
\end{align}
Connection in this case takes the form
\begin{align}
	\bar\Gamma^\alpha_{\mu\nu}  =  \Gamma^\alpha_{\mu\nu} + \frac{1}{2}\left(\delta^\alpha_\mu \phi_\nu + \delta^\alpha_\nu \phi_\mu - g_{\mu\nu} \phi^\alpha\right), \quad \phi^\mu = g^{\mu\nu}\phi_\nu,  \label{connecWeyl} 
\end{align}
with $\Gamma^\alpha_{\mu\nu}$ is the Levi-Civita connection. As one sees, in Weyl geometry $\bar\Gamma^\alpha_{\mu\nu} = \bar\Gamma^\alpha_{\nu\mu}$. Weyl connection is torsion free, but not metric preserving.

\subsection{ Lyra's geometry:}

Lyra suggested a modification of Riemannian geometry which is also a modification of Weyl geometry. The metrical concept of gauge in Weyl geometry was modified by a structure-less gauge function. The displacement vector between two neighboring points now has the components 
$\xi^\mu = x^0 dx^\mu$, where $x^0$ is a nonzero gauge function. Together with gauge function $x^0$, the coordinates $(x^\mu)$ form a reference system $(x^0; x^\mu)$. A change in reference system implies both coordinate and gauge transformation. A general transformation in Lyra is given by 

\begin{align}
	x^{\mu^\prime}  &=  x^{\mu^\prime} (x^\lambda), \quad x^{0^\prime} =  x^{0^\prime}(x^0, x^\mu), \label{lygau}
	\end{align}
	with $\partial x^0/\partial x^{0^\prime} \ne 0$ and ${\rm det} \, 	A^{\mu^\prime}_\mu \ne 0$ where $A^{\mu^\prime}_\mu = \partial x^{\mu^\prime}/\partial x^\mu.$ 
 A vector in Lyra's geometry transforms as
\begin{align}
	\xi^{\mu^\prime} &= \lambda  A^{\mu^\prime}_\mu \xi^\mu. 
\label{Lyvec0} 
\end{align}	
	If in a local reference system $(x^{0^\prime},x^{\mu^\prime})$ components of a vector remains unchanged under parallel transport, i.e.,  $\delta \xi^{\mu^\prime} = 0$, then in a general reference system $(x^0, x^\mu)$ we have 
\begin{align}	
	\delta \xi^\mu = 
	-\tilde\Gamma^\mu_{\alpha \beta} \xi^\alpha x^0 dx^\beta, \quad 
	%\label{Lyvec}\\
	\tilde\Gamma^\mu_{\alpha \beta} = \Gamma^\mu_{\alpha \beta} - \frac{1}{2} \delta^\mu_\alpha \phi_\beta,  \label{Lyvec} 
\end{align}
where we define the gauge factor \cite{Lyra}
\begin{align}
	\phi_\mu = -\frac{1}{x^0} \frac{\partial \log \lambda^2}{\partial x^\mu}, \quad \lambda = \frac{x^{0\prime}}{x^0}. \label{phidef}
\end{align}
As one sees, $\tilde\Gamma^\mu_{\alpha \beta} \ne \tilde\Gamma^\mu_{\beta\alpha}.$
A basis of tangential space in Lyra geometry is given by $\frac{1}{x^0} \frac{\partial}{\partial x^\mu}$, whereas a 1-form basis is given by $x^0 dx^\mu$. 
The metric then is introduced as follows
\begin{align}
ds^2 = g_{\mu \nu} x^0 dx^\mu x^0 dx^\nu. \label{met} 
\end{align}
The parallel transport of length in Lyra geometry is integrable, i.e.,
$\delta(g_{\mu\nu} \xi^\mu \xi^\nu) = 0$ and the connection takes the form		

\begin{align}
\bar\Gamma^\alpha_{\mu\nu}  = \frac{1}{x^0} \Gamma^\alpha_{\mu\nu} + \frac{1}{2}\left(\delta^\alpha_\mu \phi_\nu + \delta^\alpha_\nu \phi_\mu - g_{\mu\nu} \phi^\alpha\right),  \label{connecLyr} 
\end{align}
which is similar to that of Weyl geometry expect Levi-Civita connection is multiplied by the factor $1/x^0$. As one sees, $\tilde{\Gamma}^{\alpha}_{\mu\nu} = \tilde{\Gamma}^{\alpha}_{\nu\mu}$.
From \eqref{Lyvec} and \eqref{connecLyr}  it can be shown that 
\begin{equation}
	\tilde{\Gamma}^{\alpha}_{\mu\nu} = \frac{1}{x^0}\Gamma^{\alpha}_{\mu\nu} + \frac12  g^{\alpha \tau} \left( g_{\nu\tau}\phi_{\mu} -  g_{\nu\mu}\phi_{\tau} \right). \label{LyraCon}
\end{equation}
Thus, the Lyra connection is metric preserving, but not torsion free. 
The Lyra contracted curvature scalar is defined as \cite{Sen1957}

\begin{align}
	K^\lambda_{\mu\alpha\beta} &= \frac{1}{(x^0)^2} \left[\frac{\partial(x^0 \tilde \Gamma^\lambda_{\mu\beta})}{\partial x^\alpha} - \frac{\partial(x^0 \tilde \Gamma^\lambda_{\mu\alpha})}{\partial x^\beta} + x^0 \tilde \Gamma^\lambda_{\rho\alpha}
	x^0 \tilde \Gamma^\rho_{\mu\beta} - x^0 \tilde \Gamma^\lambda_{\rho\beta} x^0 \tilde \Gamma^\rho_{\mu\alpha}\right]. \label{curvature}
\end{align}

It can be shown that if the curvature is defined as above, the parallel transfer, hence the equation of motion
\begin{align}
	\frac{1}{x^0} \frac{\partial \xi^\alpha}{\partial \beta} + \tilde \Gamma^\alpha_{\nu \beta} \xi^\nu = 0, \label{em}
\end{align}
can be integrated. Einstein's field equation in Lyra's geometry in normal gauge ($x^0 = 1$) was found by Sen \cite{Sen1957}:

\begin{align}
G_\mu^\nu + \frac{3}{2} \phi_\mu \phi^\nu - \frac{3}{4}  
\delta_\mu^\nu \phi_\alpha \phi^\alpha  = \kappa
T_\mu^\nu, \label{EE}
\end{align}
where $\phi_\mu$  is the displacement vector. Here
\begin{align}
G_\mu^\nu &=	R_\mu^\nu - \frac{1}{2} \delta_\mu^\nu R,
\label{EinTen}
\end{align}
is the Einstein tensor of Riemannian geometry.

\subsection{ Spinor field}

Given the role that spinor field can play in the evolution of the
	Universe, question that naturally pops up is, if the spinor field
	can redraw the picture of evolution caused by perfect fluid and dark
	energy, is it possible to simulate perfect fluid and dark energy by
	means of a spinor field? Affirmative answer to this question was
	given in the a number of papers. We consider the spinor field
	Lagrangian given by \cite{SahaPRD2001}
\begin{align}
	L_{\rm sp} = \frac{\imath}{2} \biggl[\bp \gamma^{\mu} \nabla_{\mu}
	\psi- \nabla_{\mu} \bar \psi \gamma^{\mu} \psi \biggr] - m_{\rm sp}
	\bp \psi - \lambda_0 F, \label{lspin}
\end{align}
 where the nonlinear term $F$ describes the self-interaction of a spinor field and can be
presented as some arbitrary functions of invariants $K$ that takes one of the following values $\{I,\,J,\,I+J,\,I-J\}$ generated from the real bilinear 
forms of a spinor field such that $I = S^2 = (\bp \psi)^2,\, \&\,\, J = P^2 = (\imath \bp \gamma^5
\psi)^2$. We also consider the case $\psi = \psi(t)$  Here $\lambda_0$ is the self-coupling constant.
		
\vskip 4 mm		
The corresponding spinor field equations take the form 
	\begin{align}
	\imath\gamma^\mu \nabla_\mu \psi - m_{\rm sp} \psi - {\cD} \psi -
	\imath {\cG} \gamma^5 \psi & = 0, \label{speq1} \\
	\imath \nabla_\mu \bp \gamma^\mu +  m_{\rm sp} \bp + {\cD}\bp +
	\imath {\cG} \bp \gamma^5 & = 0. \label{speq2}
	\end{align}
where we denote ${\cD} = 2 \lambda_0 S F_K K_I$ and ${\cG} = 2 \lambda_0 P F_K 	K_J$ with  $F_K = dF/dK$, $K_I = dK/dI$ and
$K_J = dK/dJ.$ In the Lagrangian \eqref{lspin} and spinor field equations \eqref{speq1} and \eqref{speq2}
$\nabla_\mu$  is the covariant derivative of the spinor field: $\nabla_\mu \psi =
	\partial_\mu - \Omega_\mu \psi$ and $\nabla_\mu \bp = 	\partial \bp + \bp \Omega_\mu$ where $\Omega_\mu$ is the spinor affine connection defined as

\begin{align}
\Omega_\mu = \frac{1}{4} \bg_{a} \gamma^\nu \partial_\mu e^{(a)}_\nu
- \frac{1}{4} \gamma_\rho \gamma^\nu \Gamma^{\rho}_{\mu\nu}.
\label{sfc}
\end{align} 

The energy momentum tensor of the spinor
field is given by 

\begin{align}
T_{\mu}^{\,\,\,\rho} & = \frac{\imath g^{\rho\nu}}{4}  \bigl(\bp
\gamma_\mu \nabla_\nu \psi + \bp \gamma_\nu \nabla_\mu \psi -
\nabla_\mu \bar \psi \gamma_\nu \psi - \nabla_\nu \bp \gamma_\mu
\psi \bigr) \,- \delta_{\mu}^{\rho} L_{\rm sp} \nonumber\\
& = \frac{\imath}{4} g^{\rho\nu} \bigl(\bp \gamma_\mu
\partial_\nu \psi + \bp \gamma_\nu \partial_\mu \psi -
\partial_\mu \bar \psi \gamma_\nu \psi - \partial_\nu \bp \gamma_\mu
\psi \bigr)\nonumber\\
& -  \frac{\imath}{4} g^{\rho\nu} \bp \bigl(\gamma_\mu
	\Omega_\nu + \Omega_\nu \gamma_\mu + \gamma_\nu \Omega_\mu +
	\Omega_\mu \gamma_\nu\bigr)\psi
\nonumber\\
& -  \delta_{\mu}^{\rho} \lambda_0 \bigl( 2 K F_K - F(K)\bigr). \label{temsp0}
\end{align}

On account of spinor field equations \eqref{speq1} and \eqref{speq2} the spinor
field Lagrangian takes the form $L_{\rm sp} = \lambda_0 \left( 2 K
F_K - F(K)\right).$ 

\subsection{Bianchi type-VI model}

The Bianchi type-VI we take in the form

\begin{align}
	ds^2 = dt^2 - a_1^2 e^{-2mx_3} dx_1^2 - a_2^2 e^{2nx_3} dx_2^2 -
	a_3^2 dx_3^2, \label{bvi}
\end{align}
 with $a_1,\,a_2$,\, $a_3$  being the functions of time only and $m$,\,  $n$ are some arbitrary constants.
 To define the spinor affine connection $\Omega_\mu$ on account of Lyra geometry in \eqref{sfc} we should substitute $\Gamma^{\rho}_{\mu\nu}$ with $\tilde \Gamma^{\rho}_{\mu\nu}$. 
 In view of
 \begin{align}
 	g_{\mu \nu} &= e_\mu^{(a)} e_\nu^{(b)} \eta_{ab}, \label{tetmet0}
 \end{align}
 we choose the tetrad as follows:
 
 \begin{align}
 	e_0^{(0)} &= 1, \quad e_1^{(1)} = a_1 e^{-mx_3}, \quad e_2^{(2)} =
 	a_2 e^{nx_3}, \quad e_3^{(3)} = a_3. \label{tetradvi}
 \end{align}

 From $\gamma_\mu = e_\mu^{(a)} \bg_a$ one now finds
 \begin{align}
 	\gamma_0 = \bg_0, \quad  \gamma_1 = a_1 e^{-mx_3} \bg_1, \quad
 	\gamma_2 = a_2 e^{nx_3} \bg_2, \quad \gamma_3 = a_3 \bg_3.
 	\label{gbgvi}
 \end{align}
 Taking into account that in our case
 $$\bg^0 = \bg_0, \quad \bg^1 = -\bg_1, \quad \bg^2 = -\bg_2, \quad
 \bg^3 = -\bg_3,$$ one also finds
 \begin{align}
 	\gamma^0 = \bg^0, \quad  \gamma^1 = \frac{e^{mx_3}}{a_1} \bg^1,
 	\quad \gamma^2 = \frac{e^{-nx_3}}{a_2} \bg^2, \quad \gamma^3 =
 	\frac{1}{a_3} \bg^3. \label{gbgviup}
 \end{align}
 
Let us consider $\phi_\mu$ as a time-like vector field of displacement:  
 \begin{align}
 	\phi_\mu &= \{\beta(t),\,0,\,0,\,0\}, \label{phi}
 \end{align}  
  
 The connections $\tilde \Gamma^{\rho}_{\mu\nu}$ in this case are 
 
 \begin{align}
 	\tilde\G_{10}^{1} &= \tilde\G_{01}^{1} - \frac{\beta}{2} = \frac{\dot{a_1}}{a_1},
 	\quad \tilde \G_{20}^{2} = \tilde \G_{02}^{2} - \frac{\beta}{2} =
 	\frac{\dot{a_2}}{a_2},\quad 
 	 	\tilde\G_{30}^{3} = \tilde\G_{03}^{3} - \frac{\beta}{2} = \frac{\dot{a_3}}{a_3}, \nonumber\\  
 	\tilde\G_{11}^{0} &= \left(a_1 \dot{a_1}  + \frac{a_1^2 \beta}{2} \right) e^{-2mx_3},\quad \tilde\G_{22}^{0} = \left( a_2
 	\dot{a_2}  + \frac{a_2^2 \beta}{2}\right) e^{2nx_3}, \quad 
 	\tilde \G_{33}^{0} = a_3 \dot{a_3} + \frac{a_3^2 \beta}{2}, \nonumber\\
 	\tilde\G_{31}^{1} &= -m,\quad \tilde\G_{32}^{2} = n,\quad \tilde\G_{11}^{3} = \frac{m
 		a_1^2}{a_3^2} e^{-2mx_3},\quad \tilde\G_{22}^{3} = -\frac{n a_2^2}{a_3^2}
 	e^{2nx_3}. \label{Chrysvi}
 \end{align}

  The spinor affine connection in this case has the form:
\begin{subequations} 
	\label{sacbvi}
	\begin{align}
		\tilde	\Omega_0 &= -\frac{3}{8} \beta,\label{sac0}\\ 
		\tilde \Omega_1 &=  \frac{1}{2}\left[\left(\dot a_1 +  \frac{\beta a_1}{4}\right)  \bg^1 \bg^0 - m\frac{a_1}{a_3} \bg^1\bg^3\right] e^{-mx_3}, \label{G1vi}\\
		\tilde \Omega_2 &=  \frac{1}{2}\left[\left(\dot a_2 + \frac{\beta a_2}{4} \right)\bg^2\bg^0 + n\frac{a_2}{a_3} \bg^2\bg^3\right] e^{nx_3}, \label{G2vi}\\
		\tilde \Omega_3 &= \frac{1}{2} \left(\dot a_3 + \frac{\beta a_3}{4}\right)\bg^3 \bg^0. \label{G3vi}
	\end{align}
\end{subequations}

Hence the spinor field equations now read
\begin{subequations}
	\label{speqn}
	\begin{align}
		\frac{1}{x^0}\dot \psi + \frac{\dot V}{2 V} \psi  + \frac{3}{4} 
		\beta \psi - 
		\frac{m - n}{2 a_3}\bg^0 \bg^3	\psi + \imath \Phi \bg^0\psi -  {\cG} \bg^0\bg^5 \psi &= 0, \label{speq1n} \\
		\frac{1}{x^0}\dot \bp + \frac{\dot V}{2 V} \bp -            
		\frac{m - n}{2 a_3} \bp \bg^3 \bg^0 - \imath \Phi\bp \bg^0 -  {\cG} \bp \bg^0\bg^5  &= 0, \label{speq2n}
	\end{align}
\end{subequations}
where we denote $\Phi = m_{\rm sp} + {\cD}$. 
 Let us recall that in Lyra's geometry $dx^\mu$ is transforms to $x^0 dx^\mu$ and $\partial/\partial x^\mu$ to $\partial/(x^0 \partial x^\mu).$ We work in natural gauge setting $x^0 = 1$. 
Here we have defined the volume scale $V$:
\begin{align}
	V = a_1 a_2 a_3. \label{VolSc} 
\end{align}

From \eqref{speqn} one can write the equations for bilinear spinor
forms:
\begin{subequations}
	\label{inv}
	\begin{align}
		\dot S_0 + \frac{3}{4} \beta S_0   +  {\cG} A_{0}^{0} &= 0, \label{S0} \\
		\dot P_0 + \frac{3}{4} \beta P_0   -  \Phi A_{0}^{0} &= 0, \label{P0}\\
		\dot A_{0}^{0} + \frac{3}{4} \beta A^0_0  -\frac{m-n}{a_3} A_{0}^{3} +  \Phi P_0 -  {\cG}
		S_0 &= 0, \label{A00}\\
		\dot A_{0}^{3} + \frac{3}{4} \beta A^3_0  -\frac{m-n}{a_3} A_{0}^{0} &= 0, \label{A03}\\
		\dot v_{0}^{0} + \frac{3}{4} \beta v^0_0  - \frac{m-n}{a_3} v_{0}^{3} &= 0,\label{v00} \\
		\dot v_{0}^{3} + \frac{3}{4} \beta v^3_0  - \frac{m-n}{a_3} v_{0}^{0} +
		\Phi Q_{0}^{30}  +  {\cG} Q_{0}^{21} &= 0,\label{v03}\\
		\dot Q_{0}^{30} + \frac{3}{4} \beta Q^{30}_0  -  \Phi v_{0}^{3} &= 0,\label{Q030} \\
		\dot Q_{0}^{21} + \frac{3}{4} \beta Q^{21}_0   -  {\cG} v_{0}^{3} &= 0, \label{Q021}
	\end{align}
\end{subequations}
where we denote $S_0 = S V,\, P_0 = P V,\, A_0^\mu = A_\mu V,\,
v_0^\mu = v^\mu V,\, Q_0^{\mu \nu} = Q^{\mu \nu} V$.  
After some manipulations from \eqref{inv} one can obtain the following relation between invariants of the  spinor field: 
\begin{subequations}
	\label{inv0}
	\begin{align}
		(S_{0})^{2} + (P_{0})^{2} + (A_{0}^{0})^{2} - (A_{0}^{3})^{2} &= C_1 \exp[-(3/2) \int \beta(t) dt], \label{inv01}\\
		(Q_{0}^{30})^{2} + (Q_{0}^{21})^{2} + (v_{0}^{3})^{2} -
		(v_{0}^{0})^{2} &= C_2 \exp[-(3/2) \int \beta(t) dt],  \label{inv02}
	\end{align}
\end{subequations}
where $C_1$ and $C_2$ are integration constants. From \eqref{inv} it can also be shown that 
 \begin{align}
	S &= \frac{C_0}{V}{\exp{\left[-\frac{3}{4}\int \beta (t) dt\right]}}, \quad
	K = \frac{C_0^2}{V^2}{\exp{\left[-\frac{3}{2}\int \beta(t) dt\right]}}, \label{K}
\end{align}
where $C_0$ is a constant. 
\vskip 4mm

On account of \eqref{sacbvi} 
in BVI cosmological model the spinor field possesses the following components of the energy momentum tensor \cite{Saha2016,Saha2018}:
\begin{subequations} 
	\label{emt} 
\begin{align}
	T_0^0 &= m_{\rm sp} S + \lambda_0 F := \varepsilon, \label{T00}\\
	T_1^1 &= T_2^2 = T_3^3 = \lambda_0 \left(F - 2 K F_K\right) := -p, \label{Tii}\\
	T_1^0 &= -\frac{n\, e^{-m x_3}}{4}  \frac{a_1}{a_3} \, A^2 , \label{T10}\\
	T_2^0 &= -\frac{m\, e^{n x_3}}{4} \frac{a_2}{a_3} \,A^1, \label{T20} \\
	T_2^1 &= \frac{e^{(m +n) x_3}}{4} \frac{a_2}{a_1} \left[\left(\frac{\dot
		a_1}{a_1} - \frac{\dot a_2}{a_2}\right) A^3
	- \frac{m + n}{a_3}A^0\right] , \label{T12}\\
	T_3^1 &= \frac{e^{m x_3}}{4} \frac{a_3}{a_1}
	\left(\frac{\dot a_3}{a_3} - \frac{\dot a_1}{a_1}\right) A^2 \label{T21}\\
	T_3^2 &=\frac{e^{-n x_3}}{4} \frac{a_3}{a_2}
	\left(\frac{\dot a_2}{a_2} - \frac{\dot a_3}{a_3}\right)A^1, \label{T32}
\end{align}
\end{subequations}	
where $\varepsilon$ is the energy density and $p$ is the pressure.

\subsection{Einstein equation}

In view of \eqref{emt} the Einstein's equations corresponding to \eqref{bvi} we rewrite in the form \cite{Saha2016,Saha2018}:

	\begin{subequations}
		\label{EEs}
		\begin{align}
			\frac{\ddot a_2}{a_2} + \frac{\ddot a_3}{a_3} +
			\frac{\dot a_2}{a_2}\frac{\dot a_3}{a_3}  &= \kappa \lambda_0 \left(F - 2 K F_K\right)  + \frac{3}{2} \beta^2 + \frac{n^2}{a_3^2}, \label{EE11}\\
			\frac{\ddot a_3}{a_3} + \frac{\ddot a_1}{a_1} +
			\frac{\dot a_3}{a_3}\frac{\dot a_1}{a_1} &= \kappa \lambda_0\left(F - 2 K F_K\right)  + \frac{3}{2} \beta^2 + \frac{m^2}{a_3^2}, \label{EE22}\\
			\frac{\ddot a_1}{a_1} + \frac{\ddot a_2}{a_2} +
			\frac{\dot a_1}{a_1}\frac{\dot a_2}{a_2} &= \kappa \lambda_0\left(F - 2 K F_K\right) + \frac{3}{2} \beta^2  - \frac{m n}{a_3^2}, \label{EE33}\\
			\frac{\dot a_1}{a_1}\frac{\dot a_2}{a_2} + \frac{\dot
				a_2}{a_2}\frac{\dot a_3}{a_3} + \frac{\dot a_3}{a_3}\frac{\dot
				a_1}{a_1}  &=   \kappa\left(m_{\rm
				sp} S + \lambda_0 F\right) - \frac{3}{2} \beta^2 + \frac{m^2 - m n + n^2}{a_3^2} , \label{EE00}\\
				\left(m - n\right) \frac{\dot a_3}{a_3} - m \frac{\dot a_1}{a_1} + n
				\frac{\dot a_2}{a_2}  &= 0, \label{EE03}\\
			0 &= -\frac{n\, e^{-m x_3}}{4}  \frac{a_1}{a_3} \, A^2 , \label{EE10}\\
			0 &= -\frac{m\, e^{n x_3}}{4} \frac{a_2}{a_3} \,A^1, \label{EE20} \\
			0 &=\left[\left(\frac{\dot a_1}{a_1} - \frac{\dot a_2}{a_2}\right) A^3
			- \frac{m + n}{a_3}A^0\right] , \label{EE12}\\
			0 &=\left(\frac{\dot a_3}{a_3} - \frac{\dot a_1}{a_1}\right) A^2 \label{EE21}\\
			0 &=\frac{e^{-n x_3}}{4} \frac{a_3}{a_2}
			\left(\frac{\dot a_2}{a_2} - \frac{\dot a_3}{a_3}\right)A^1. \label{EE32}	
		\end{align}
	\end{subequations}
	
From \eqref{EE03} one dully finds the relation between the metric functions
\begin{align}
a_3 &= X_0 \left(a_1^m/a_2^n\right)^{1/(m-n)},\quad X_0 = {\rm const.}, \label{abcrel}
\end{align}
whereas, from \eqref{EE10} - \eqref{EE32} on account of \eqref{A03} leads to
\begin{align}
A^2 = 0, \quad  A^1 = 0, \quad A_{0}^{3} =
\left(a_1/a_2\right)^{(m-n)/(m + n)} \, \exp{[-(3/4 \int \beta (t) dt)]}.
\label{A3a12}
\end{align}

In view of Bianchi identity $	G_{\mu;\nu}^\nu \equiv 0$ from \eqref{EE} we find 

\begin{align}
 \frac{3}{2} (\phi_\mu \phi^\nu)_{;\nu} - \frac{3}{4}  
	\delta_\mu^\nu (\phi_\alpha \phi^\alpha)_{;\nu}  &= \kappa
	T_{\mu;\nu}^\nu. \label{EED}
\end{align}
It should be noted that this result depends on how we define covariant derivative. In Riemannian geometry the connection is symmetric in two lower indices, so it does not matter how one writes the indices. But in Lyra geometry when it is no longer symmetric, the notation matters a lot. For example if one defines covariant derivative as in \cite{Hooft} 
\begin{align}
	\nabla_\alpha A_\mu^\nu &= A_{\mu ; \alpha}^\nu = \partial_\alpha A_\mu^\nu - \Gamma^\lambda_{\alpha \mu} A_\lambda^\nu + \Gamma^\nu_{\alpha \lambda} 
	A_\mu^\lambda , \label{Hooft} 
\end{align}
the energy conserves, i.e., in that case $T^\nu_{\mu;\nu} = 0$. But if one uses standard definition \cite{Landau} 
\begin{align}
	\nabla_\alpha A_\mu^\nu &=  A_{\mu ; \alpha}^\nu = \partial_\alpha A_\mu^\nu - \Gamma^\lambda_{\mu\alpha} A_\lambda^\nu + \Gamma^\nu_{\lambda \alpha} A_\mu^\lambda, \label{Landau} 
\end{align}  
we find that energy in this case no longer conserves and on account of spinor field equations we obtain 
\begin{align}
	T^\nu_{0;\nu} = \frac{3}{4} \beta \left(\varepsilon + p\right). \label{NEP} 
\end{align}
Inserting \eqref{NEP} into \eqref{EED} we find the equation for $\beta$: 

\begin{align}
	\dot \beta + \left(\frac{\dot a_1}{a_1} + \frac{\dot a_2}{a_2} + \frac{\dot a_3}{a_3} + \frac{3}{2}\beta\right) \beta 
	&= \frac{\kappa}{2} (m_{\rm sp} S + 2 \lambda_0 K F_K).  \label{beta}
\end{align}

Note that in earlier works we have obtained metric functions and invariants of spinor field in terms of volume scale $V$. Equations for $V$ was derived Einstein equation:
\begin{align}
	\ddot V &= 2 \frac{m^2 - mn + n^2}{a_3^3} V + \frac{3}{2} \left[m_{\rm sp} S + 2 \left(F - K F_K\right)\right]  \label{VDef} 
\end{align} 
As one sees, the equation for volume scale does not contain the Lyra parameter $\beta$ explicitly, but through spinor field invariants it influences the evolution of space-time. In what follows we solve the equations for spinor and gravitational fields with Lyra geometry numerically.

\section{Numerical solutions}

In this section we solve the Einstein equation numerically. For that we rewrite Eqns. \eqref{EE11} - \eqref{EE33} together with \eqref{beta} in the following way introducing directional Hubble parameter: 

\begin{subequations}
	\label{sys01}
	\begin{align}
		\dot a_1 &= H_1 a_1, \label{a1} \\
			\dot a_2 &= H_2 a_2, \label{a2} \\
				\dot a_3 &= H_3 a_3, \label{a3} \\
				\dot H_1 &= - H_1^2 + \frac{1}{2}\left(H_2 H_3 - H_1 H_2 - H_3 H_1 \right) \nonumber\\
				&+ \frac{1}{2} \left( \kappa \lambda_0 \left(F - 2 K F_K\right)  + \frac{3}{2} \beta^2  \right)  + \frac{m^2 - n^2 - mn}{2 a_3^2}, \label{H1}\\
					\dot H_2 &= - H_2^2 + \frac{1}{2}\left(H_3 H_1 - H_1 H_2 - H_2 H_3 \right) \nonumber\\  &+ \frac{1}{2} \left( \kappa \lambda_0 \left(F - 2 K F_K\right)  + \frac{3}{2} \beta^2  \right) 
					+ \frac{n^2 - m^2 - mn}{2 a_3^2}, \label{H2}\\
						\dot H_3 &= - H_3^2 + \frac{1}{2}\left(H_1 H_2 - H_2 H_3 - H_3 H_1 \right) \nonumber\\ &+ \frac{1}{2} \left( \kappa \lambda_0 \left(F - 2 K F_K\right)  + \frac{3}{2} \beta^2  \right) + \frac{m^2 + n^2 + mn}{2 a_3^2}, \label{H3}\\
			\dot \beta &= - \left( H_1 + H_2 + H_3 + \frac{3}{2}\beta\right) \beta + \frac{\kappa}{2} (m_{\rm sp} S + 2 \lambda_0 K F_K).  \label{beta1}			
	\end{align}
	\end{subequations}
	Beside this system from \eqref{EE00} and \eqref{EE03} we have 
	
	\begin{align}
	H_1 H_2  + H_2 H_3  + H_3 H_1  &=   \kappa\left(m_{\rm
		sp} S + \lambda_0 F\right) - \frac{3}{2} \beta^2 + \frac{m^2 - m n + n^2}{a_3^2}, \label{EE00a}
	\end{align}
and 
\begin{align}
	 H_3 &=  \frac{m H_1 - n H_2}{(m - n)}.
 \label{H312}
\end{align}
Moreover, we have the relation \eqref{abcrel} between metric functions. Hence it is sufficient to numerically solve equations for $a_1,\,a_2,\,H_1$, $H_2$ and $\beta (t)$. On account of \eqref{abcrel} and \eqref{H312} we now rewrite the system \eqref{sys01}. 

\begin{subequations}
	\label{sys01z}
	\begin{align}
		\dot a_1 &= H_1 a_1, \label{a10} \\
		\dot a_2 &= H_2 a_2, \label{a20} \\
			\dot H_1 &= \frac{1}{2(m-n)}\left[ (2n - 3m) H_1^2 - n H_2^2 + 2 n H_1 H_2\right] \nonumber\\ 
			&+ \frac{1}{2} \left( \kappa \lambda_0 \left(F - 2 K F_K\right)  + \frac{3}{2} \beta^2 \right)  
		+ \frac{m^2 - n^2 - mn}{2 a_3^2}, \label{H10}\\
		\dot H_2 &= \frac{1}{2(m-n)}\left[ m H_1^2 + (3n - 2m) H_2^2 - 2 m H_1 H_2\right] \nonumber\\  
		&+ \frac{1}{2} \left( \kappa \lambda_0 \left(F - 2 K F_K\right)  + \frac{3}{2} \beta^2 \right) 
		+ \frac{n^2 - m^2 - mn}{2 a_3^2}, \label{H20}\\
		\dot \beta &= - \frac{1}{m-n} \left[ (2m -n) H_1 + (m - 2n) H_2 \right] \beta + \frac{3}{2}\beta^2  + \frac{\kappa}{2} (m_{\rm sp} S + 2 \lambda_0 K F_K).  \label{beta2}	
		\end{align}
\end{subequations} 
The eqn. \eqref{EE00a} now can be written as

\begin{align}
	\frac{m}{m-n} H_1^2  + 2 H_1 H_2  - \frac{n}{m-n} H_2^2  &= \left[ \kappa\left(m_{\rm
		sp} S + \lambda_0 F\right) - \frac{3}{2} \beta^2 + \frac{m^2 - m n + n^2}{a_3^2}\right]. \label{EE00a1}
\end{align}
Here $V = X_0 (a_1^{2m + n} a_2^m)^{1/(m-n)}$ and $a_3 = X_0 \left(\frac{a_1^m}{a_2^n}\right)^{1/(m-n)}.$

To solve this system we have to write nonlinear term of spinor field explicitly. Recall that setting $K= I = S^2$ from \eqref{S0} one dully finds $S = C_s e^{-(3/4) \int \beta(t)dt}$, whereas in case of mass-less spinor field for $K = J = P^2$ from \eqref{P0} one finds  $S = C_p e^{-(3/4) \int \beta(t)dt}$. Analogically from \eqref{S0} and \eqref{P0} for mass-less spinor field one obtains  $K = I \pm J = C_k e^{-(3/2) \int \beta(t)dt}$. 

\begin{figure}
	\centering
	\includegraphics[width=11 cm]{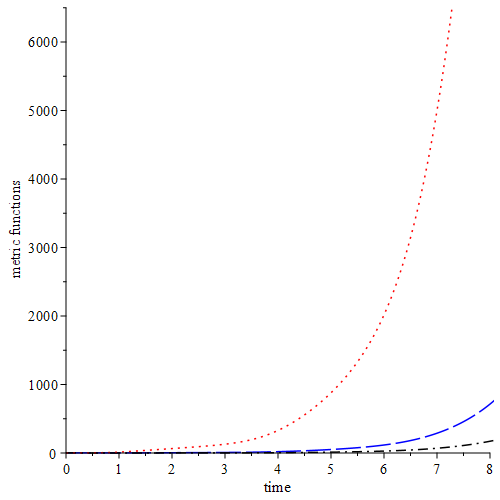}
	\caption{\label{Fig1}Evolution of metric functions $a_1(t)$ (blue long dash),  $a_2(t)$ (red dash-dot) and $a_3(t)$ (black solid) with Lyra geometry when spinor field nonlinearity simulates  modified Chaplygin gas.}
\end{figure}

\begin{figure}
	\centering
	\includegraphics[width=11 cm]{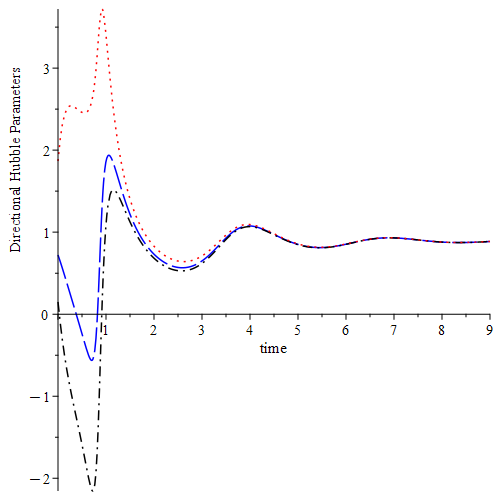}
	\caption{\label{Fig2}Evolution of directional Hubble parameters $H_1(t)$ (blue long dash),  $H_2(t)$ (red dash-dot) and $H_3(t)$ (black solid) with Lyra geometry when spinor field nonlinearity simulates  modified Chaplygin gas.}
\end{figure}

\begin{figure}
	\centering
	\includegraphics[width=11 cm]{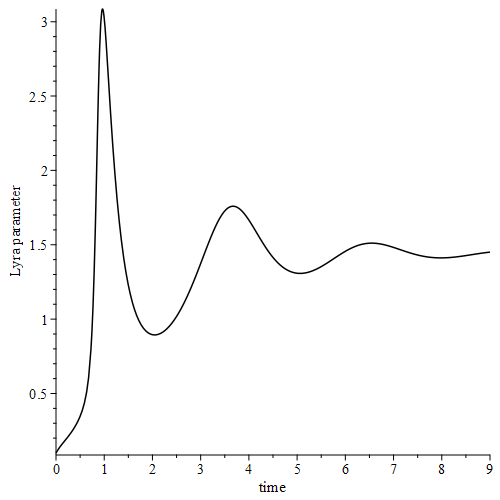}
	\caption{\label{Fig3}Evolution of Lyra parameter $\beta(t)$ when spinor field nonlinearity simulates  modified Chaplygin gas.}
\end{figure}

We consider spinor field nonlinearity that describes a modified Chaplygin gas:
\begin{align}
	F = \left(\frac{A}{1 + w} + \lambda_1 K^{(1+w)(1+\alpha)/2} \right)^{1/(1+\alpha)}, \label{mch} 
\end{align}
with $A > 0$, $0 \le \alpha \le 1$. Here $\lambda_1$ is a integration constant.
It can be shown that the spinor field nonlinearity in this case satisfies 
\begin{align}
	p = W \varepsilon - A/\varepsilon^\alpha. \label{mchdef} 
\end{align}
For numerical solution we have set $\kappa = 1$, $\lambda = 1$, $\lambda_1 = 1$, $w = 1/2$, $\alpha = 1/3$, $A = 1$, $m = 3$ and $n = 2$. As the initial values we set $a_1 (0) = 0.97$, $a_2(0) = 1.03$, $h_2 (0) = 0.15$, $\beta (0) = 0.1$, $R (0) = 0$ and $h_1 (0)$ is calculated from \eqref{EE00a1} that gives $h_1 (0) = \{-0.822,\,0.722\}$. In our calculations we use $h_1 (0) = 0.722$. The metric functions and directional Hubble parameters are displayed graphically in Fig. \ref{Fig1} and Fig. \ref{Fig2}, respectively. In Fig. \ref{Fig3} we have demonstrated the evolution of Lyra parameter $\beta (t)$.      

\section{Concluding remarks}

Within the framework of Bianchi-type VI space-time with Lyra geometry we have studied the role of spinor field in the evolution of the Universe. It is found that in this case energy-momentum tensor of spinor field does not preserve.  
The invariants of spinor field have exponential dependence of the parameter of Lyra geometry. Though the non-diagonal components of EMT do not vanish and still impose different kinds of restrictions on space-time geometry and spinor field, still the introduction of Lyra geometry significantly influences the whole process.


\begin{thebibliography}{9999}
	\bibitem{Weyl} H. Weyl, {\it Gravitation and Electricity}, Preuss. Akad. Wiss. Berlin, 465 (1918) 	
	
	\bibitem{Lyra} G. Lyra,  Math. Z. {\bf 54}, 52 (1951)
	
	\bibitem{Scheibe} E. Scheibe, Math. Z. {\bf 57}, 65 (1952) 
		
		\bibitem{Sen1957} D. K. Sen, Z. Physik. {\bf 149}, 311 (1957)
		
					
				\bibitem{Halford} Halford J. Math. Phys. {\bf 13}, 1699 (1972)
				
					\bibitem{Sen1971} D.K. Sen and K.A. Dunn, J. Math. Phys. {\bf 12}, 578 (1971) 
					
					
		\bibitem{Sen1972} D.K. Sen and J.R. Vanstone, J. Math. Phys. {\bf 13}, 990 (1972)
		
		\bibitem{Soleng} H.H. Soleng, Class. Quantum Grav. {\bf 5}, 1489 (1988) 
		
		\bibitem{Beesham} A. Beesham, Aust. J. Phys. {\bf 41}, 833 (1988)  
		
		\bibitem{Jahromi} A.S. Jahromi and H. Moradpour, Int. J. Mod. Phys. D {\bf 27} 1850024 (2018)  
		
			\bibitem{Bakry} M.A. Bakry, Astrophys. Space Sci. {\bf 367}, 35 (2022)
			
				\bibitem{Shchigolev} V.K. Shchigolev and D.N. Bezbatko, Grav. $\&$ Cosmology, {\bf 24}(2), 161 (2018)
		\bibitem{SahaPRD2001} B. Saha, Phys. Rev. D {\bf 64}, 123501 (2001)
		
		\bibitem{Saha1997GRG} Saha B. and Shikin G.N.  Gen. Relat. Gravit. {\bf 29}(9), 1099 (1997)
		
		\bibitem{Saha1997JMP} Saha B. and Shikin G.N. J. Math. Phys. {\bf 38}(10), 5305 (1997)
		
			
		\bibitem{Greene}  Armend$\acute a$riz-Pic$\acute o$n C., Greene P.B.
		Gen. Relat. Grav. {\bf  35}(9), 1637 (2003)
		
		\bibitem{ELKO}  Fabbri L. Phys. Rev. D.  {\bf 85}, 047502 (2012)
		
		\bibitem{PopPRD}  Pop{\l}awski N.J. Phys. Rev. D. {\bf  85}, 107502 (2012)
		
		
		\bibitem{Saha2016} B. Saha, Eur. Phys. J. Plus {\bf 131} 170 (2016)
		
		\bibitem{Saha2018} B. Saha, Phys. Part. Nucl. {\bf 49}(2), 146 (2018)
		
		\bibitem{Casana}  R. Casana, C. A. M. de Melo, B. M. Pimentel, Astrophys. Space Sci. {\bf 305}, 125 (2006)
		
		\bibitem{SahaJMP} Bijan Saha, J. Math. Phys. {\bf 66}(10), 102501 (2025)
					
			\bibitem{Hooft} Gerard ’t Hooft, {\it Introduction to General Relativity} (2012) 
			
			\bibitem{Landau} L.D. Landau and E.M. Lifshitz, {\it The Classical Theory of Field}, Pergamon Press Ltd. (1971)   
		
	\end{thebibliography}
	\end{document}